\documentclass[11pt]{article}
\usepackage{geometry}                
\usepackage{graphicx}
\usepackage{epsfig}
\usepackage{amssymb,amsfonts,amsmath,amsthm}
\usepackage{mathrsfs}
\usepackage{epstopdf}
\usepackage{hyperref,color}
\usepackage{natbib}
\usepackage{setspace}
\usepackage{enumitem}

\pdfoutput=1

\title{The philosophy of causal set theory}
\author{Christian W\"uthrich\thanks{I thank Fay Dowker and Sumati Surya for discussions and feedback on earlier versions of this article. This work was supported financially by the John Templeton Foundation (the views expressed are those of the author, not necessarily those of the sponsors).}}
\date{8 August 2023}

\begin{document}
\maketitle

\begin{center}
{\small Written for the section on causal set theory edited by Fay Dowker, Rafael Sorkin, and Sumati Surya to be included in Cosimo Bambi, Leonardo Modesto, and Ilya Shapiro (eds.), \textit{Handbook of Quantum Gravity}, to be published with Springer Nature, Singapore.}
\end{center}

\begin{abstract}\noindent
This article presents the most interesting philosophical issues as they arise in causal set theory. The first concerns the apparent disappearance of spacetime at the fundamental level. It shows how the looming empirical incoherence is averted if we adopt spacetime functionalism. Second, classical sequential growth dynamics rekindles hope for a fundamental passage of physical time compatible with relativistic physics. The article argues that this hope is faint at best, as a block view offers the most natural interpretation of dynamical causal set theory. Third, causal set theory admits a very natural structuralist interpretation, enabling a fruitful interaction between debates in philosophy of science concerning structural realism and the metaphysics of causal sets.
\end{abstract}


\section{Introduction}
\label{sec:intro}

A central task---perhaps {\em the} central task---of natural philosophy, as understood more broadly than physics, is to explain the manifest image of the world on the basis of the scientific image, in Wilfried Sellars's memorable terms. In other words, natural philosophy is to deliver an account of how the world can manifest itself to us as it does if it is structured as our best science tells us. This may sound like a trivial commonplace not worth pausing over, but as fundamental physics gets ever more removed from our direct experience of the world, the task, while remaining eminently important, turns increasingly delicate. 

As we move in our quest for a quantum theory of gravity beyond the well-trodden standard model of particle physics and general relativity (GR), the task becomes both more critical and more fragile than ever before. It becomes more critical because the apparent non-spatiotemporality of the fundamental structures postulated in approaches to quantum gravity threatens the empirical coherence of these approaches. In order to avert this threat, it needs to be shown how these structures yield, at the appropriate scales, a framework for the familiar spatiotemporal world. It becomes more fragile because the currently leading approaches to quantum gravity tend to postulate fundamental structures rather different from the relativistic spacetimes to which they are supposed to give rise. This increased gap requires hard technical and conceptual work to be bridged, and, alas, success is not guaranteed. 

The degree to which the structures of quantum gravity will be non-spatiotemporal is an interesting and involved question, as is the potential philosophical fallout. As argued at length in the forthcoming monograph by \citet{hugwut}, major approaches to quantum gravity such as string theory, loop quantum gravity, and causal set theory all postulate fundamental structures which turn out to be non-spatial or non-temporal in significant ways. We will return to the issue of how spatial and how temporal causal sets really are in \S\ref{sec:emerge} below, but let us suppose, for the sake of argument, that the fundamental structures are indeed non-spatiotemporal in significant ways. The problem of empirical incoherence, then, is that the following three propositions cannot all be maintained:\footnote{The presentation of the problem of empirical incoherence here is close to that of \citet[\S6.1.1]{Yat:19}. An earlier formulation of the problem can be found in \citet{hugwut13}.}

\begin{enumerate}
\item A fundamental theory in physics is empirically coherent only if it delivers empirical predictions.
\item Empirical predictions in physics are formulated in terms of local beables, and so presuppose their existence. 
\item The fundamental ontology of physics is non-spatiotemporal.
\end{enumerate}
Crucially, it is assumed that physics presupposes what \citet[234]{bel87} dubbed `local beables', i.e., things which are real (exist) or are at least candidates for being---hence `be-ables'---and which are associated with some determinate region of spacetime---hence `local'. In other words, empirical predictions in physics assert something about physical entities (fields, particles, pulsars, more directly photographic plates, Geiger counters, pointers, computer readouts, etc), which they take to be localizable in space and time, and their states. Crucially, local beables can only exist if space and time (or better {\em spacetime}) exist. Given the first two propositions above, it is a necessary condition for the empirical coherence of a fundamental theory in physics that there be spacetime. However, if the third proposition is true, then that necessary condition can apparently not be satisfied, leaving the theory empirically incoherent. 

In essence, the problem of empirical incoherence represents one important aspect of the central task of natural philosophy, as it links the fundamental structures of quantum gravity ultimately to how the world manifests itself to us, which is, among other things, evidently spatiotemporal. We will consider the disappearance and emergence of spacetime in causal set theory in \S\ref{sec:emerge}.

The following section, \S\ref{sec:time}, will discuss another philosophical issue in causal set theory, which has recently risen to prominence: the interpretation of the dynamics in causal set theory and its implications for the philosophy of time. The problem of empirical incoherence and the emergence of spacetime can be considered a {\em foundational} problem in the context of causal set theory (and other approaches to quantum gravity), i.e., a possibly philosophical problem which arises in the context of developing and interpreting our best physical theories. In other words, in foundational problems, we start from physics and become in some sense `philosophical' while our ultimate interests remain in physics. In contrast, in {\em philosophical} problems, we start out from originally philosophical questions and turn to physics, hoping to find at least partial answers to our ultimately philosophical questions. Let it be noted that this distinction is at best a first approximation, as it depends on there being a principled distinction between physics and philosophy, a distinction I believe is at best gradual and approximate when it comes to topics such as those treated in this article. 

As \S\ref{sec:time} is concerned with the nature of time and whether time is essentially dynamical or not and turns to causal set theory to see which side in this debate garners support from quantum gravity, we are here faced with a philosophical question in this sense. In this section, we will mainly consider how the classical sequential growth dynamics proposed by \citet{Rideout1999} ought to be interpreted and whether or not it thus supports a metaphysics of `becoming', of temporal dynamism. The main conclusion of this section will be that the `block universe' view still seems preferable in light of fundamental physics. However, although causal set theory takes up some prior arguments in favour of dynamism from the context of GR, it also adds some genuinely novel and interesting twists to the debate.

Before reaching the conclusions in \S\ref{sec:conc}, \S\ref{sec:structure} considers the metaphysics of causal sets more generally, and in particular the nature of the basal events. As it turns out, structuralism offers a very natural interpretation of causal sets. As a structuralist interpretation rejects the idea that fundamental entities have intrinsic natures beyond their structural roles, this raises the issue of how to deal with elements of so-called `non-Hegelian sets', i.e., basal events with exactly the same relational profile: if the identity of an event is exhausted by its relational profile, it seems as if there cannot be distinct events with the same relational profile. 

Apart from probing deep questions about the metaphysics of causal sets, these interpretive considerations straightforwardly enrich the important debate in philosophy of science surrounding the formulation and tenability of structural realism. While I will argue that the structuralist interpretation provides the natural template for causal set theory (as it possibly does for other physical theories), I will caution against taking this result to deliver strong support for a thoroughgoing form of structural realism.

\section{Emergence}
\label{sec:emerge}

As I have stated in \S\ref{sec:intro}, part of the wider task of fundamental physics is to tie its theories back to the world as we experience it. For theories in quantum gravity, this task involves showing how something closely approximated by relativistic spacetimes emerges from the fundamental structures postulated. This is not a novel situation in physics: whenever a successor theory is proposed, one of its central duties is to deliver the old theory, or something sufficiently close to it, in an appropriate limit or approximation. The task, and particularly what would count as its successful completion, is not given in terribly precise terms and remains open to renegotiation in the scientific proceedings. Ultimately, it is the peers in the scientific community who will, collectively, decide over success or failure. It is clear that delivering on this task is a conditio sine qua non for the acceptance of novel theories in fundamental physics.   

\subsection{The basics of causal set theory}
\label{ssec:basics}

Consequently, it is one of the central problems in causal set theory to show how the causal sets it postulates are approximated by relativistic spacetimes. Perhaps this task would be straightforward if causal sets were clearly spatiotemporal. After all, there is a reasonable hope that they are, as they are so closely modelled on what is taken to be the essential features of relativistic spacetimes. Causal set theory, as a research program, starts out from a series of results in classical GR which culminated in what is known as `Malament's theorem' \citep{mal77}. Roughly put, Malament's theorem states that for causally sufficiently well behaved spacetimes, their full geometry can be reconstructed from the causal relations among events plus a conformal factor. One of the elegant aspects of causal set theory is that if we replace the continuum of relativistic spacetime by a discrete structure, then this conformal factor is fixed in a rather natural way: the cardinality of a subset of a causal set offers a natural measure of its `size'. In this way, causal set theory departs from a beautifully simple starting point: the discrete fundamental structure is ordered by a relation of causal precedence. 

Fundamentally, a causal set is thus a discrete partial order.\footnote{Causal set theory was first proposed in \citet{bomeal87}; for reviews, see \citet{Dowker2013,Sorkin2009b,Surya2019}; for a philosophical review, see \citet[chapter 3]{hugwut}. Reference to chapters in this handbook.} First, the discreteness is imposed a priori. This has some technical advantages as some notorious divergences which show up in a continuum theory are thus avoided. However, the main motivation, I take it, is that the fundamental structure which will give rise to relativistic spacetimes is assumed to be a discrete structure because this is what one could expect, perhaps on the basis that in quantum theories, many physical observables have discrete spectra. The proof of the pudding is in the eating, which for a theory in fundamental physics means that its basic posits, whatever their original motivation, must receive confirmation in the usual ways of science: the theory is empirically correct and usefully predictive, and it enjoys theoretical advantages such as explanatory power or simplicity over its empirically equivalent rivals. Theories of quantum gravity to date remain, of course, far from being confirmed.

Second, events are partially ordered by causal precedence in Minkowski spacetime, but this is not generally true in GR. Given that there are pairs of spacelike-related events in relativistic spacetimes which do not stand in a relation of causal precedence, demanding that the order be total would clearly be too strong and would violate a central insight of relativistic physics. However, the demand of events being ordered partially is still too strong, as GR permits models of spacetimes which contain closed timelike curves. In those spacetimes, the relation of causal precedence is no longer antisymmetric, and so not a partial order.\footnote{A binary relation $R$ is {\em antisymmetric} over a set $C$ just in case $\forall x, y \in C$, if $xRy$ and $yRx$, then $x = y$. Clearly, if there are closed timelike curves, then a spacetime event $a$ could both precede and be preceded by an event $b$ in that spacetime. Since $a$ and $b$ are distinct, antisymmetry is violated.} Again, that the fundamental order is partial is a stipulation of the theory which may be vindicated a posteriori by the theory's success; but it does seem to rule out as unphysical relativistic spacetimes which are not causally sufficiently well behaved that the causal ordering is partial.\footnote{However, this appearance may be false, as is argued in \citet{wut21}.}

Third, it should be noted that the theory as stated thus far is not a quantum theory, as of course is the goal. How a quantum theory of gravity based on classical causal set theory will look like is not clear.\footnote{For a review, see \citet{Surya2019}, particularly \S6.3.} Perhaps the quantum aspects of the theory will be confined to its dynamics (see the next point); or we might theorize that the state of the world is generically a superposition of causal sets; or its quantum nature is expressed in yet other ways. For the purposes of the present article, we will restrict ourselves to the classical theory as it stands now. 

Finally, the mere stipulation that the fundamental structure be a discrete partial order turns out to be far too weak. In fact, there is a rigorous sense in which almost all causal sets will be `pathological' insofar as they will not deliver anything near a useful model of the cosmos. As it turns out, almost all sufficiently large discrete partial orders consist of only three highly connected `layers' or `generations' of basal elements \citep{klerot75}. If considered cosmological models, almost all sufficiently large causal sets would represent worlds with highly non-local causal connections and which `last' for a mere three Planck times, during which they double in size from the first `moment' to the second, and then halve in size as they transition to the final `moment'. It is thus clear that additional conditions must be imposed in order to arrive at a theory whose postulated structures more generically qualify as promising candidates to model fundamental physical reality. There are potentially many ways in which a useful restriction to viable models could be accomplished. Although it may be attractive to focus on additional conditions with a prima facie physical plausibility, the conditions imposed will ultimately (and a posteriori) be judged by the success of the resulting theory. Causal set theorists typically assume that these conditions should specify a reasonable dynamics which in some sense governs the `growth' of a causal set---a sense which will be studied in \S\ref{sec:time} below; only causal sets which could have come into being by a process compatible with the stipulated dynamics will be deemed physically possible.

\subsection{The non-spatiotemporality of causal sets}
\label{ssec:nonspt}

In spite of the fact that the resulting causal sets have been closely modelled on relativistic spacetimes, there are some significant differences. The most salient departure from relativistic models is that causal sets are not obviously spatiotemporal. Consequently, the problem of empirical incoherence lurks.

How are causal sets less than fully spatiotemporal? They are obviously discrete structures, but their non-spatiotemporality runs deeper than that. Although the fundamental ordering relation is a relation of {\em causal} precedence, it shares features one would expect to hold of a {\em temporal} precedence relation. If we rule out temporal loops and assume the transitivity of temporal precedence, as seems intuitive, then temporal precedence is also antisymmetric, just as causal precedence was assumed to be. In relativistic theories, temporal precedence, however one concretely defines it then at most orders events partially, again just as does causal precedence. It seems as if causal precedence is structurally similar to a minimal form of relativistic temporal precedence with neither metric relations such as durations nor any `flow' or passage of time.

The extent to which one takes this distinction between causation and time seriously depends on one's metaphysical position regarding the relation between causation and time. In fact, \citet[136n]{dowker20} takes the fundamental relation in causal set theory to be one of temporal, not causal, precedence and asserts that it would thus be more appropriate that the approach be called `temporal set theory'. Against this stance, one could argue that the causal structure of relativistic spacetimes, on which the relation is directly modelled, is primarily causal, and at best derivatively temporal. After all, this structure tracks the causal connectibility of events by light signals. It can be shown \citep[\S3.1.1]{hugwut} that important results such as Malament's theorem are closely intertwined with the history of causal theories of time. These theories invert the usual metaphysical hierarchy between time and causation and assume that causation is the more fundamental of the two and that, consequently, time derives from causation, not vice versa.\footnote{See \citet{baronlebihan2023} for a recent version of such a causal theory of spacetime.} If this is right, then the fundamental relation in causal set theory is one of causal precedence and time will only emerge from the causal structure. 

Space is much more clearly absent from causal set. Without going into an analysis of the essence of space, space is usually so called because it has a certain topological and geometrical structure which allows us to speak of (typically unary, binary, ternary, or quaternary) spatial relations of being nearby, far away, between, collinear, orthogonal, parallel, three-dimensional, etc. One might argue that many of these relations track a folk (or perhaps Euclidean) notion of space and are absent in GR also, and that our task is merely that of recovering recovering relativistic spacetime, implicitly assuming that Euclidean space (or Newtonian spacetime) approximates relativistic spacetime in certain states. 

This point is perfectly valid, but as it turns out, there is a sense in which what ought to be taken as `space' in a causal set has none of these structures. In fact, it has {\em no structure at all}. Spatial structure only emerges as we consider ever larger and sufficiently well-behaved causal sets. In order to see this, let us first identify what should reasonably be taken as `space' in a causal set. In a relativistic spacetime, one way to construct space (at a time) is to introduce a foliation of spacetime, i.e., a partition of the four-dimensional Lorentzian manifold into three-dimensional spacelike hypersurfaces which are ordered by the values of a real-valued smooth function with nowhere vanishing, timelike gradient. The partition is then interpreted as a slicing of spacetime into space at subsequent moments in time. 

In a causal set, lacking much of this structure, a foliation of the kind we seek would still partition the causal set into subsets which are ordered and labeled by a sequence of integers. These subsets would then represent space at a moment in time, at least if an additional condition is satisfied: their elements have to be pairwise `incomparable', i.e., none of their elements can causally precede, or be causally preceded by, another element of the same subset. In technical terms, this means that subsets must be `antichains'. If this condition of being spatial were violated, then causal precedence could be `instantaneous'---and the fundamental relation should certainly not be considered `temporal' in any way. Furthermore, the antichains should be {\em inextendible} in the sense that any element of the causal set not in the antichain causally precedes or is causally preceded by an element of the antichain. For finite partially ordered sets, it can be shown that such a partition into antichains always exists \citep[55]{bri97}. Just as for relativistic spacetimes, however, these foliations will in general be highly non-unique. 

One can think of these antichains as representing `space' (at a `time'), and of the partition as a sequence of `nows'. The trouble, it turns out, is that antichains are by definition entirely structureless sets. `Space' as we have identified it has no topological or geometric structure at all. Space, in causal set theory, is altogether absent: causal sets seem to have no spatial structure. 

However, if causal sets give rise to something well approximated by relativistic spacetimes at some scale, then we will have to be able to reconstruct something from causal sets which resembles the spatial structure of relativistic spacetimes. In more general terms, we will have to find a way to extract geometric and topological information from fundamental causal sets in order to relate them to spacetimes.

\subsection{Spacetime functionalism and the emergence of spacetime}
\label{ssec:func}

As the previous section has shown, it seems clear that causal sets are less than fully spatiotemporal. In order to avoid the problem of empirical incoherence, and more generally to retrieve the manifest image of the world from the scientific one, it is necessary to show how relativistic spacetimes `emerge' from causal sets. More specifically, this means that it needs to be established how relativistic spacetimes are excellent approximations to causal sets at certain scales or in certain regimes. 

As I have argued with co-authors on several occasions \citep{hugwut13,lamwut18,lamwut20,hugwut}, {\em functionalism} offers the right template to understand the relationship between fundamental structures as postulated and described in theories of quantum gravity and relativistic spacetimes and in that represents a key tool to discharge the central task outlined at the outset. Generally speaking, functionalism identifies an entity not by its internal constitution, but instead by its `functions' or the roles it plays. Functionalism about spacetime, then, claims that what makes something spacetime is that it `plays the spacetime role'. In particular, spacetime functionalism charges us with two subtasks. 

First, spacetime, or spacetime properties, are `functionalized'. This means that we characterize spacetime in terms of its functions or of the roles it plays, e.g.\ in our theories in physics or, ultimately, in giving a scientific explanation of our phenomenology of the world. Without offering a comprehensive answer here, it is clear that spacetime fulfils roles such as determining the relative localization of physical entities, ordering events in time, metrical relations such as spatial distances or temporal durations, and the like. An important point to be noted regarding the first step is that functionalism does not insist that any particular entity, such as `spacetime' itself, exists: it remains silent on whether (relativistic) spacetime exists. Instead, any kind of fundamental substances or properties are permitted, as long as they play the appropriate functional roles. In fact, these roles can be multiply realized. The general slogan of spacetime functionalism, as captured in the title of \citet{lamwut18}, is `spacetime is as spacetime does'.

Second, once the functions of spacetime are specified, we provide an explanation of how the fundamental entities or properties can execute these functional roles. Given the multiple realizability permitted in the first step, the fundamental entities which fill the roles of spacetime may themselves be quite different from relativistic spacetime. However different they turn out to be, however, an explanation of how they nevertheless fulfil the relevant functional roles of spacetime must be given. For instance, this means that it must be shown how the fundamental structures deliver (relative) localization of entities, the ordering of events in space and time, and have spatial distances and temporal durations, and so on, at the emergent level. 

If this functionalist agenda is successfully completed for a research program in quantum gravity, then the threat of empirical incoherence is thwarted: as the spacetime features necessary for empirical confirmation are shown to be available at the relevant scale in the approach at stake, empirical confirmation can then proceed in the usual way {\em even if the fundamental ontology of the theory diverges significantly from that of relativistic spacetimes}. 

In light of the fact that approaches to quantum gravity remain active research programs with the as of yet unfulfilled ambition of delivering a complete and concretely worked out (and empirically confirmed) theory of quantum gravity, the second step of the functionalist agenda can only be sketched. Although causal set theory is no exception and remains a work in progress, its advocates have started to outline in some detail how a functionalist program might be implemented in causal set theory, although of course not under this name. Let us consider some work in this direction.\footnote{For more details, see \citet[\S4]{lamwut18} and \citet[chapter 4]{hugwut}.}

First---and this is vital to the functionalist strategy---, causal set theory needs not to recover the whole continuum or manifold structure of GR. However, some causal sets ought to be `manifoldlike' in order to show that they are well approximated by relativistic spacetime described by manifolds of reasonably low dimensionality, and with Lorentz signature and non-pathological causal structure. By showing that causal sets satisfying the dynamical laws are `manifoldlike', one can thus establish that they satisfy a necessary condition to perform the relevant spacetime functions. This involves showing that there exists an embedding of a causal set into a spacetime such that the causal relations are preserved, the mapping distributes the elements of the causal set sufficiently uniformly into the manifold, and the spacetime has no non-trivial structure at scales below the mean point spacing. 

Suppose that a sufficiently large causal set is manifoldlike in this sense and is well approximated by a physically reasonable spacetime. The next task is then to show that the approximating spacetime is `approximately unique'.\footnote{That this is indeed the case is the famous `Hauptvermutung'.} This condition is imposed in order to ensure the uniqueness of the emerging spacetime and so that one and the same causal set cannot play inconsistent spacetime roles. It turns out that both a rigorous formulation of this condition as well as proof of its satisfaction are scientifically hard problems that have so far defied complete control. 

The next, and final, task of the functionalist research program is to show how causal sets can fill the roles of spatiotemporal localization, spatial distance, temporal duration, topology, etc. This job requires the construction of concrete means of extracting this kind of information from causal sets. Although much remains to be done, physicists have published substantive work in this direction. Recall from \S\ref{ssec:nonspt} that causal sets have no spatial structure at all. Establishing how causal sets can give rise to spatial geometry and spatial topology thus becomes a central problem to be addressed in the functionalist program. Indeed, physicists have worked to define spatial topology \citep{Major2006,Major2007} in terms native to the fundamental causal set, and similarly with spatial structure and distance \citep{ridwal09a,ridwal09b}. These efforts aim to deliver on the second functionalist task of showing how the fundamental structures can play spacetime roles and thus constitute paradigm examples of functionalist work. 

If this second step is completed, then it is fully established that causal sets are suitable to perform the relevant roles of spacetime. On the functionalist paradigm, completing the two steps of the program is all there needs to be done to show how spacetime emerges in causal set theory and thus to avert the threat of empirical incoherence. 

There are at least two ways in which this conclusion can be resisted. The first challenges the original verdict according to which the fundamental structures are non-spatiotemporal, the second accepts the verdict, but puts doubt on the success of the functionalist program. Roughly put, this tracks a dilemma posed by \citet{Yat:19} with which he struck spacetime functionalism: either our theory of the fundamental structures is relevantly isomorphic to the theory of spacetime it replaces, or it is not.\footnote{To be more precise, Yates strikes a particular form of spacetime functionalism with this dilemma, so-called `realizer spacetime functionalism'.} If it is, then we should certainly expect that the functional reduction succeeds in deriving spatiotemporal structures from the fundamental ones; but then the fundamental structures were spatiotemporal after all, rendering spacetime functionalism otiose. If it is not, then it is hard to see how spacetime functionalism could connect the non-spatiotemporal fundamental theory to higher-level spacetime theories, delivering its impotence. Either way, spacetime functionalism fails to do the work which it was advertised to do. 

Let me address this dilemma before returning to other objections. It is clear that spacetime functionalism had a problem if it failed to connect the two levels, as it was indeed built to do precisely that. So it better avoid the second horn of the dilemma. However, success, as on the first horn, does not imply that the fundamental ontology was spatiotemporal and that therefore, there was no point in introducing functionalism. In fact, we have seen above how causal sets are at least not directly spatial, indeed spatially structureless, and recovering spatial structure is rather elaborate and highly indirect, as the literature on this cited above testifies. Substantive work is necessary to establish the connection. Moreover, not all causal sets will be sufficiently `well behaved' to give rise to non-degenerate spatiotemporal structure, further supporting the fact that causal sets are not directly and automatically spatiotemporal. In general, the situation in quantum gravity is far murkier than is suggested by Yates's dilemma and the fundamental structures are not so easily comparable to relativistic spacetimes. In fact, this connection is even more tenuous in some other approaches to quantum gravity than it already is in causal set theory. 

Returning to the two ways of resisting the general conclusion of this section, by denying non-spatiotemporality or by rejecting functionalism, one finds traces of both in the literature. As for the first camp, \citet{leblin19} argue that many approaches to quantum gravity, including causal set theory, postulate fundamental structures, which are spatiotemporal, or at least include a fundamental asymmetry between space and time. The latter, rather weak, claim according to which there are clear signs in causal set theory of a distinction between space and time certainly seems true, but too weak to change anything that was said above. In order to challenge our conclusions, the much stronger claim that causal sets are rather directly spatiotemporal would have to be supported. Even if we were to interpret the causal relations as relations of temporal precedence, as we have seen above in \citet[136n]{dowker20}, this would not change anything about the fact that direct spatial structure is absent in causal sets. Given the significant differences between causal sets and relativistic spacetimes, this first family of objections does not seem to cut very deeply into our conclusions.

As for the second type of objections, there are of course standard objections against functionalism in the context of the philosophy of mind, from where the position has been borrowed.\footnote{\citet[\S5]{lev21} is a standard reference for these objections to functionalism as a program in the philosophy of mind. See \citet{lamwut18} for responses to analogous objections to spacetime functionalism.} Without rehashing these general objections here, let us focus on two related criticisms. In the slightly different context of wave function realism, \citet{ney15}'s `macro-object objection' raises concerns about spacetime functionalism being unable to account for the constitution of macroscopic objects, such as four-dimensional objects of our experience and perhaps spacetime itself. Just like a hologram may capture some of the features of the object it depicts but nevertheless falls short of fully constituting that object, spacetime functionalism cannot deliver a satisfactory account of the constitution of such objects. However, as explicated in \citet[\S3]{lamwut18}, the macro-object objection relies on a very robust and ultimately spatiotemporal notion of constitution, which can be avoided altogether by endorsing a non-spatiotemporal mereology \citep{lebbar22}. Furthermore, taking spacetime functionalism seriously implies that, faithful to its slogan `spacetime is as spacetime does', if there is a nature or an essence to spacetime, then it is exhausted by the functions it plays in our theories in physics, or perhaps for understanding human experience more generally. Requiring that the constituents of spacetime have a true, intrinsically spatiotemporal (but completely ineffable) nature is just to beg the question against spacetime functionalism. If we can show that each and every empirically relevant role of spacetime can be played by our fundamental structures, there remains nothing else to show. 

Relatedly, \citet{le2021spacetime} distinguishes between ``hard'' and  ``easy problems'' of spacetime emergence, in a loose analogy to the problem of consciousness in the philosophy of mind. He invokes the cognitive dissonance aroused by the inconceivability of a non-spatiotemporal world as evidence for an explanatory gap, which spacetime functionalism cannot close. Le Bihan (S374) insists that a physical `what it is like', such as the spatiotemporal ordering of events through experience or the metricity of their spatiotemporal relation stands in need of explanation. Without further going into the details here,\footnote{Those can be found in \citet[\S2.4]{hugwut}.} and while acknowledging the existence of a cognitive dissonance about the emergence of spacetime, the program of spacetime functionalism, if successfully executed, closes any explanatory gap that needs closing. 

In sum, the (partial) non-spatiotemporality of the fundamental ontology of theories of quantum gravity, including causal set theory, turns out to open deep and fruitful philosophical questions concerning the nature of spacetime. I have argued that these questions are best addressed by adopting spacetime functionalism.

\section{Philosophy of time}
\label{sec:time}

Leaving these foundational issues behind, we turn to philosophical issues as they arise in the context of causal set theory. Perhaps the most obvious philosophical implications of causal set theory are to be found in the philosophy of time, at least judging by the number of authors and papers devoted to this topic: \citet{Arageorgis12}, \citet{Butterfield2007}, \citet{Dowker2003,dowker14,dowker20}, \citet{Earman2008}, \citet{hug14}, \citet{Sorkin2006,Sorkin2007}, \citet{wutcal15}. At the center of the debate resides the question whether causal set theory, unlike much of the rest of contemporary fundamental physics, supports a notion of `becoming' against the standard eternalist orthodoxy. This section follows the main arguments on both sides of the debate. 

\subsection{The debate in philosophy of time}
\label{ssec:debate}

In philosophy of time, two main camps face one another. On the one hand, we have those who favour a metaphysics of time which places the apparent dynamical aspects of time on centre stage, the sense that things `become', that there is a `passage of time', that time `flows'. Views in this family are chiefly motivated by intuition based on the apparent phenomenology of temporality. They inevitably give a fundamental role in their metaphysics to an ever changing, updating, and dynamically advancing `present'. {\em Presentists} consider present events and objects the only ones to really exist, with past ones having passed away and future ones yet to become. For presentism, the sum total of existence thus only contains present entities. {\em Growing block theorists} also admit past entities as genuinely existing, with the sum total of what exists being presented by an ever growing block to which new slivers of existence are continually being added as they become. The present is awarded a special status in that it represents the cusp of the growing block, the advancing front of becoming. 

In contrast to these dynamical metaphysical views, we find, on the other hand, those who eliminate this dynamical aspect from their fundamental metaphysics. Fundamentally, there is `being', but no `becoming'. Becoming, to the extent to which it is an objective feature of reality, emerges at some scale, perhaps from interactions between fundamental physics with the cognitive apparatus of perceiving agents such as human beings. Views in this camp are often, though not invariably, motivated by contemporary fundamental physics. {\em Eternalism} is the view that the present does not play a special role and that, consequently, present entities do not enjoy a special ontological status. Eternalism is sometimes described as the view according to which past, present, and future entities all exist on a par. This characterization is problematic in that it still seems to presuppose a distinction between past, present, and future when eternalism denies that there is, objectively and fundamentally, any such distinction to be had. With this distinction gone, eternalism then accepts that the dynamical features of time which depended on it cannot be fundamental. The sum total of existence according to eternalists contains what presentists would call past, present, and future entities. 

There is a danger that the debate is trivialized with both sides agreeing on obvious facts. It turns out that it is surprisingly difficult to articulate presentism as a substantive metaphysical position such that it is neither trivially true (because it just asserts that nothing exists---now---that is not present) nor obviously false (because it commits to nothing existing---at some time or other---that is not present).\footnote{This concern has been articulated many times over, although in somewhat different forms, for example in \citealt{lom99}, \citealt{cal00}, \citealt{mey05}, and \citealt{sav06}.} All hands agree that dinosaurs existed in the past, but that they no longer exist now. But presentism, and the entire debate between the distinct metaphysical theses in play here, cannot be captured adequately if we think of `existing' as implicitly demanding a temporal locution: either existence is existence {\em now} or it is existence {\em at some time or other}. We need a concept of existence {\em simpliciter} free of any such temporal implication. 

In spite of what one sometimes reads in the literature, eternalists also think that dinosaurs do not exist now. They think that dinosaurs exist {\em simpliciter}, i.e., that they are part of the sum total of existence; but it is simply not the case that eternalists think that everything, including past and future entities, exists {\em now}.\footnote{For just one prominent example of a problematic characterization of eternalism, see \citet[\S6]{emery20}.} The locution `now' is simply inadmissible in fundamental discourse and can at best be an indexical which functions like `here' and `I'. Indexicals like `here' and `I' and, the eternalist would add, `now', do not figure in fundamental descriptions of our world. Consequently, any aspect of our metaphysics which depends on a fundamental present cannot be part of the fundamental description. In general, we are again led to the conclusion that `existence' cannot imply existence at a particular time for the position (and hence the debate) to be meaningful.

Perhaps a useful way of thinking about the debate is to recognize that there are obvious sub- and superset relations among the sum total of existence according to the three positions \citep{wut12c}. For instance, from the presentist perspective, the eternalist is committed to a strict superset of entities compared to their commitment, and from the eternalist perspective, the growing block's sum total of existence is a strict subset of theirs. More could be said in an attempt to make this more rigorous, but I trust the idea is sufficiently clear for us to proceed on this understanding.

\subsection{The situation before causal set theory}
\label{ssec:before}

Before the advent of relativistic physics, when physics was thought to describe what happens in a Newtonian setting, any of our three metaphysical positions could straightforwardly be combined with our best physics. Although Newtonian physics did not invoke a distinction between past, present, and future, it is easily seen as being compatible with it. However, once we move to relativistic physics, this symmetry is broken in favour of eternalism. The dynamicist theories of presentism and the growing block run into the problem of depending on a fundamental distinction between what is present and what is not in a relativistic setting where simultaneity between spacelike related events is relative to a frame of reference. Consequently, in relativity there is no global, objective, frame-independent and thus absolute notion of present available. And it seems as if existence ought to be a global, objective, and frame-independent affair. Therefore, existence cannot depend on a notion of present, as presentism and the growing block theory demand. It thus seems as if eternalism is the only game in town once we accept relativity.\footnote{Although the argument is fairly direct, it seems as if a version of it is first found in \citet{rie66} and \citet{put67}. It has been repeated many times since (for instance in \citealt{skl81} or \citealt{sau02}). For a more general assessment of the prospects of presentism in modern physics, see \citealt{wut13}.} 

The above argument implicitly depends on the `present' being defined by an equivalence relation of co-presentness which is naturally satisfied by events on a spacelike hyperplane. A foliation of spacetime into a (totally) ordered set of spacelike hypersurfaces is unique in Newtonian (or neo-Newtonian) spacetime, but is relative to the frame of reference and so non-unique in special relativity's Minkowski spacetime. However, as \citet{Stein1991} has proved, there exists a (unique) frame-independent absolute and non-trivial relation of co-presentness which may underwrite an objective notion of becoming. In particular, given an event in Minkowski spacetime as vantage point, we can identify all events on its past lightcone as being `co-present' with it. Such a relation of co-presentness permits the definition of versions of presentism (only an event and events on its past lightcone exist) or the growing block theory (which also admits events inside the event's past lightcone). 

However, saving presentism or the growing block from the relativity of simultaneity comes at a price. First, since the new relation of co-presentness is non-transitive and antisymmetric and so clearly not an equivalence relation, the usual intuitions invoked to motivate presentism seem to become more removed from what is supposed to be the form of the present. Defending such a lightcone presentism would generally require giving up rather natural intuitions about time. For example, it would have to be denied that the causal ordering of events along null geodesics, such as of the creation and detection of a photon, implies a temporal ordering---lightlike-related events like this would be co-present. More of interest for present purposes is a second point: although absolute in the sense of frame-independence, the present is relative to a particular reference {\em event}, the `given' event. To be sure, presentism in Newtonian spacetime also picked one of the spacelike hypersurfaces as the present. But this happened not at the expense of the other hypersurfaces: those would all take turns in sequentially becoming the `present'. In this way, every event in Newtonian spacetime would exist at some point in time. However, on the present proposal, it is not clear how (a) a particular event is chosen as the `given' event, and (b) how the dynamical updating is supposed to work. What exists appears to depend on the arbitrary choice of an event as reference, unless something is said about the dynamics of the `present'. 

A natural way to generalize the view emerging from Stein's proposal, and suggested in \citealt{CliftonHogarth}, is to relativize becoming to a given (infinite) worldline and then state that the dynamical sequence of nows is given by the totally ordered past lightcones of events along that worldline. In this way, every event in Minkowski spacetime will be said to exist at some point in time (namely when it is `swept over' by the past lightcone of events on the worldline). The worldline needs to extend from past to future infinity, and ought to be the worldline of a possible observer. Call this `worldline-dependent becoming' and note that it is certainly available already in Minkowski spacetime. 

Worldline-dependent becoming is {\em objective} in that it only relies on the geometry of Minkowski spacetime. Furthermore, it is {\em absolute} in that it is frame-independent, i.e., is based only on Lorentz-invariant structures. However, it is relative in that it depends on a particular given worldline. Thus, even though it does not privilege a particular frame of reference, it sanctions one particular worldline or observer. I will call this feature of worldline-dependent becoming {\em local}. 

Apart from this form of localism, there are other unattractive features of worldline-dependent becoming. `Being co-present' is no longer an equivalence relation, as both transitivity and symmetry no longer hold. Although this loss of equivalence may not be fatal for the view, it has clearly unpalatable metaphysical consequences: if we tie existence to co-presentness, then an entity $a$ may exist for another entity $b$ while $b$ does not exist as far as $a$ is concerned. Furthermore, events which on some intuitive notion of global time lie in the distant past will be co-present if they are spatially sufficiently far removed. This is certainly odd, but since in Minkowski spacetime nothing answers to this intuitive notion of global time, there may simply not be enough structure to make the worry stick. 

These unwelcome consequences can be mitigated if we either base our metaphysics on some non-Lorentz invariant structure or accept a relativization of existence to frames of reference. The former would amount to an unscientific hypostatization of an undetectable structure. Perhaps surprisingly, the latter has been defended in the literature: \citet{fine2005} defends `fragmentalism', i.e., the view which accepts that our commitment to presentism and special relativity, forces us to accept that existence simpliciter is relativized to frames of reference such that different inertial observers will in general disagree as to what exists, not merely as to what is simultaneous. 

In light of special relativity, advocates of dynamical theories such as presentism or the growing block view thus face the following dilemma: either their metaphysics answers to their initial motivation and explanatory requests or is compatible with the structure of Minkowski spacetime, but not both \citep{cal00,wut13}. They must thus either give up their original ambition or go against very well established physics. 

Is this uncomfortable position ameliorated as we go beyond special relativity to more fundamental theories? The dilemma stays essentially the same as we move to GR, even though the debate is enriched by two new factors, pulling in opposite directions. The good news for the presentist first: as it turns out, in spacetimes of an important family of models in GR, there exists a physically privileged foliation. These spacetimes admit an objective cosmic time, thus grounding an objective and in principle observable distinction of events into past, present, and future. The Friedmann-Lema\^itre-Robertson-Walker (FLRW) spacetimes, which form the backbone of the cosmological standard model, belong to this family. Given that cosmologists have good evidence for thinking that these models describe the spacetime structure of our actual world with a surprisingly high accuracy at sufficiently large scales and at sufficiently late times, the presentist might be tempted to draw (premature!) hope. Although the FLRW spacetimes appear to correctly capture the large-scale structure of spacetime, they assume a perfectly uniform distribution of matter-energy across the universe. The rather significant local deviations from such a global average distribution thwarts the local validity of a partition into past, present, and future---most crassly in black holes. Moreover, it is not clear how the global average distribution of the universe's matter-energy content could be causally connected to the intuitions which drive presentism. 

Although the presentist may find ways to finesse these difficulties, they will have to content themselves with a metaphysics of time, which can at best be contingently true: many models of GR, and thus many ways in which GR deems the world could have been, do not admit any foliation at all. This is the bad news: rather than an embarrassment of riches as in Minkowski spacetime and other relativistic spacetimes where foliations into spacelike hypersurfaces were highly non-unique, in these cases, there is no way at all to partition spacetime into past, present, and future. If we restrict ourselves to naturalism, it seems as if the presentist has two main strategies available: either they forgo the idea of global present in favour of a more local notion, or else they make the case that those unfoliable spacetimes are, although formally models of GR, not physically reasonable possibilities. The latter option, while prima facie reasonable, will involve the stubborn challenge to deliver quite general reasons why unfoliable spacetimes should not be considered physically reasonable, lest we have to articulate specific reasons spacetime by spacetime. We will return to the former strategy in the next subsection.

\subsection{Philosophy of time in causal set theory}
\label{ssec:timecst}

As we have seen in the previous subsection, the central principles of relativistic physics seriously limit the scope for theories of time which include a fundamental notion of `becoming', although they fall short of ruling them out altogether. Causal set theory promises to broaden that scope and to brighten the prospects of a dynamical metaphysics of time. The basis of that promise is that on its standard formulation, which includes a dynamics such as `classical sequential growth' dynamics \citep{Rideout1999}, causal set theory can be interpreted to postulate an ``active process of growth in which `things really happen' '' \citep{Sorkin2006}, a `birthing' of elements of a causal set, without violating any of the central tenets of relativity, such as general covariance, the general principle ultimately responsible for the difficulties for `becoming' in relativistic physics. In order for such an active process of becoming to be compatible with relativistic principles, a global form of becoming is replaced by a local version, in line with the first strategy in the last paragraph of the previous subsection.

Following the structure in \citet{wutcal15}, let us consider the fate of dynamical theories of time first at the kinematical level of causal set theory, before turning to classical sequential growth dynamics. A growing causal set closely resembles a discrete version of a growing spacetime block, and so I will often just speak of the growing block theory. However, a presentist position can easily be gleaned from dynamical causal set theory: the present consists of all and only the maximal elements of a dynamically growing causal set. In this way, each dynamical addition of another element, which can only happen to the future of previously added elements, `updates' the present, i.e., the set of maximal elements at that `moment'. Although the growing block view may give us a more natural template for interpreting the dynamically growing causal set due to the asymmetry between the presence of a `birthing' process and the absence of an `annihilating' process, it is easy enough to modify the interpretation to suit presentist needs. The reader is invited to keep this in mind when I will speak only of the growing block in the remainder of the section. 

Let's start at the kinematic level. The advocate of a dynamical theory will seek to foliate causal sets into slices of subsequent `nows'. We can identify a maximal antichain, i.e., a maximal set of events which are pairwise incomparable with respect to the fundamental relation $\prec$ of causal precedence, as the universe at a `moment of time'. Any finite causal set admits a partition of the entire structure into maximal antichains and is thus foliable. For infinite causal sets, the question of foliability involves some subtleties which preclude a fully general answer. Suffice it to say that the past-finite causal sets grown by classical sequential dynamics are foliable in our sense. We can thus safely assume that the physically relevant class of causal sets admits a foliation of the entire structure into a sequence of subsequent presents. 

There are some parallels to the situation in GR, and a few notable differences between the continuum relativistic spacetimes and the discrete causal sets. First, the parallels. In both cases, foliations are highly non-unique, although there is a sense in which the cardinality of the non-uniqueness is higher in the continuum case of GR. The partition of the base structure into a sequence of subsequent presents is extraneous to the physical theory in both cases (just as SR or GR did not single out a particular foliation, causal set theory also does not). That this addition is extraneous can also be seen from the fact that in both cases, the partitions are not invariant under automorphisms of the base structure. 

As for the differences, there are good reasons to think that the physically relevant causal sets afford such a partition, whereas in GR there exist numerous non-foliable spacetimes. If we can state good reasons for excluding those, then this difference will disappear. As for a definite difference, a maximal antichain in causal set theory, which is supposed to represent the universe at a `moment of time', has absolutely no intrinsic structure. This is radically different from the situation in GR, where we will find a rich induced spatial geometry and topology in a space-like hypersurface. Finally, there is no analogue of Stein's special-relativistic theorem in causal set theory, due to the existence of `non-Hegelian subsets', to be discussed in \S\ref{sec:structure} below. The failure of Stein's theorem means that we can generally hope for more freedom in trying to identify a relation appropriate for becoming. However, as \citep[\S3]{wutcal15} argue, the defender of a dynamical theory cannot find much traction in this failure, which means that the original dilemma for this position between acceptance of the physics and maintaining an attractive position still stands.

Still at the level of kinematics, if we seek a `becoming' interpretation beyond the narrow confines of imposing a foliation into something like space-like hypersurfaces or maximal antichains, we can straightforwardly identify causal set theory analogues of worldline or lightcone becoming: a worldline is a chain of events connected by $\prec$, and the past lightcone of an event is the set of all events which causally precede that event in terms of $\prec$. This straightforward identification arguably reinforces the original dilemma \citep{wutcal15}. And of course, the alternative block interpretation is equally available in causal set theory as it is in relativity.  

The kinematics of causal set theory are thus most naturally interpreted to be devoid of real becoming, just as standard GR. Claims of real becoming in causal set theory are all based on the dynamics, which is added to the kinematics in order to restrict the models of the theory to physically reasonable ones. The standard (though classical) dynamics is a law of sequential growth (more specifically, of `generalized percolation') such that a causal set grows by a sequential addition of new elements to the causal future of existing ones, where the elements to which the new element is causally related is a matter of probability \citep{Rideout1999}. Although not yet quantum, classical sequential growth dynamics is a natural and useful stepping stone toward a path-integral formulation of an eventual quantum theory of causal sets. In this dynamics, `Becoming' appears to be embodied in this sequential addition of new elements, a process which is interpreted by \citet[024002-2]{Rideout1999} to be constitutive of time:
\begin{quote}
The phenomenological passage of time is taken to be a manifestation of this continuing growth of the causet. Thus, we do not think of the process as happening `in time' but rather as `constituting time' [...]
\end{quote}
Before I proceed to discuss the prospects of becoming in a fully dynamical causal set theory, let me emphasize that a `block universe' interpretation remains very much a live option, as \citet{hug14} also remarks.\footnote{This is his first interpretive option (page 16); the second augments the kinematical causal structure with a gauge-invariant dynamics, to be discussed below.} Under this interpretation, the dynamics just offers a space of possible full histories with a probability measure defined on them. It is thus clear that dynamical causal set theory remains fully compatible with a metaphysics of non-dynamical being, a block universe without becoming.

Let us consider the viability of a metaphysics of becoming in causal set theory augmented by a dynamics of classical sequential growth. The most obvious path to such an interpretation is by turning the `now' into a localized, observer-dependent matter, i.e., a form of worldline (or lightcone) becoming: individual observers experience local becoming as they inch up on their worldlines towards the future.\footnote{That local, asynchronous becoming is closely analogous to worldline or lightcone becoming is also noted by \citet{Arageorgis12}.} In the words of \citet{Sorkin2007}, rather than ``super observers'', we have an ``asynchronous multiplicity of `nows' ''. 

Although there thus exist analogues of localized, `asynchronous' becoming in GR in the form of worldline or lightcone becoming which are just as covariant as asynchronous becoming in dynamical causal set theory, \citet{dowker14,dowker20} wants to drive a wedge between asynchronous becoming in causal set theory and its analogues in GR. What is missing in GR for the analogy to hold, it seems, is any valid reason to think of spacetime events as not merely existing, or ``having happened'' or ``will have happened'', but instead as ``happening'' as the result of a dynamical process of ``occurrence''. In GR, according to Dowker, events are just there without ever being born, or undergoing (or having undergone) a process of occurrence. Although the {\em result} of the birthing occurrences of spacetime events in dynamical causal set theory is the same---that the event is there, exists---the path that leads to the result is essentially different: whereas in the case of birthing occurrences, we have true becoming, in the block universe, we find just static `being'. 

I believe that the analogy between asynchronous becoming in causal set theory and lightcone becoming in GR is much tighter than Dowker seems to think, for two reasons.\footnote{\citet{hug14} argues that unless the background `time' relative to which events are born can somehow be shown to be physical, i.e., not mere gauge, the dynamics is fully analogous to what we find in GR and so not hospitable to a substantive notion of passage. Showing background time to be physical is Huggett's second option.} First, it should be noted that lightcone becoming in GR does not require the introduction of global forms of time and thus is fully covariant in any way one might demand. It just needs the local causal structure, very much like asynchronous becoming in causal set theory. 

Second, it seems difficult to maintain that the asynchronous becoming in causal set theory---the `birthing' of events---is a physical process. If it were, then it would be rather unusual for such a process to not occur in (or constitute) {\em physical} time. Let me explain. The discrete form of general covariance imposed on the growth dynamics in causal set theory---necessary to keep the theory properly relativistic---and the consequent absence of any facts of the matter which of two unrelated (and so `spacelike-related') events `occurred' or was `born' first, there is no physical background time in which the births occur. This was precisely why there was no global time and that the resulting becoming is asynchronous. As a consequence, in both GR and causal set theory, there is no room for a global notion of becoming, but clearly scope for a localized, asynchronous form of it, rendering the analogy rather tight.

However, \citet{wutcal15} have identified two novel ways in which becoming can become more global in causal set theory than it ever could in GR, both of which are ultimately due to the discreteness of the fundamental structure. 

The first is that there appears to be a global physical fact about the size of the universe at any stage of the birthing of events. Although the order in which unrelated events occur must remain indeterminate due to the required covariance, the cardinality of the causal set at any stage of the sequential growth is an objective, global, gauge-invariant fact.\footnote{For the related concept of `covtree', see \citet{Zalel23} in this Handbook.} First, the causal set has zero events, then one, then two, then three, etc. At any stage $n$ of the growth process, we can thus affirm that the causal set consists of $n$ elements and so has a determinate size, we cannot, in general, assert which events have occurred by stage $n$. Consequently, there is a sense in which we have ontological indeterminacy as to which events have already occurred by a given stage, but without any indeterminacy in the cardinality of the structure representing the sum total of existence at that stage. 

The second novel feature is truly exotic, and I am not aware of any other context in which something similar can be found. One might expect, on the basis of the first feature, that no particular event in a future-infinite causal set is ever going to enjoy determinate existence until the asymptotic future, as it were, when the infinite growth process has been completed. Once the growth of the causal set has been completed, and all births have happened (which will of course not be the case at any finite stage), then all events will snap into determinate existence from their prior indefinite state. Even for future-finite causal sets, there is the analogous worry that events will not come into determinate being until the growth has been completed, which will occur after a finite number of steps. 

However, this is in general not the case, at least for future-infinite causal sets growing by transitive percolation. In order to see this, let us introduce the concept of a `post': a {\em post} is an event that is either causally preceded by or causally precedes any other event in the causal set. A post can be interpreted as an event where the universe undergoes a transition from a sharply contracting to an expanding phase, perhaps a `big bang' of sorts. As it turns out \citep{Rideout1999}, causal sets grown by transitive percolation in general have many such posts. Suppose that the event born at stage $n$ of such a causal set is a post. In this case, all events in the causal past of the post and the post itself will snap into determinate existence, leaving no ontological indeterminacy at that stage. At the next stage, however, there will again be ontological indeterminacy, unless the next event is also a post.\footnote{See \citet[\S4, particularly figure 3]{wutcal15} for a fuller explanation and a figure illustrating the point.} 

Unfortunately, it is unclear whether the supposition that a particular event {\em is} a post can be legitimately posited. The problem is the following: {\em whether or not} a given event is a post remains itself indeterminate until the causal set has fully grown. As long as the growth process is ongoing, it remains possible that an event is added at some later stage which is causally unrelated to an event we might have thought of as a post; if that happens, then the original event is of course no longer a post and the ontological determinacy of its past cannot be assumed. 

Note just how exotic this new form of becoming is. While it is of course quite natural on a dynamic metaphysics of time to consider the future indeterminate, we have here a literal sense in which the past is indeterminate also, or at least what would be a natural analogue of the past in a causal set. In GR, there is a sense in which the past changes from indeterminate to determinate in lightcone becoming as the lightcone grows and encompasses larger parts of the `past'. However, in causal set theory, all of the `past' is generally indeterminate until the end, when the entire causal set becomes determinate at once. 

In conclusion, it thus seems as if all events in a dynamically growing causal set, including `past' ones, remain ontologically indeterminate until the growth process is completed. At that stage, finite or not, we have the full causal set, and the resulting ontology is indistinguishable from one based on the block universe metaphysics. Thus, we either accept a block interpretation, or else we purchase a foreign form of becoming in the coin of a rather complete ontological indeterminacy. Events may become, but only indeterminately so. It should thus be clear that becoming in dynamical causal set theory assumes a novel, exotic form.

\section{Structuralism}
\label{sec:structure}

In philosophy of science, `structural realism' is offered to the realist as a means to evade the strictures of the pessimistic metainduction \citep{Worrall1989}. The pessimistic metainduction strikes the realist with the repeated embarrassment of having committed to obsolete ontologies whenever a scientific theory gets replaced with a successor theory. The structuralist analysis of the pessimistic metainduction identifies the problem in pre-structuralist realism in its assumption that we ought to commit to an ontology of `things'. If instead of reading off of theories a surface ontology of objects, we committed to their underlying structures, the pessimistic argument is avoided because the structures the structural realist commits to survive scientific revolutions. At the same time, it is these same structures which are responsible for the predictive success of theories, and not their surface ontologies. Consequently, {\em structural realism} commits to the structures described by the mathematical formulations of the pertinent theories, and only to these structures, and thus hopes to maintain the explanatory and predictive power of the theory. Structural realism as a general strategy to respond to the pessimistic metainduction offers a wholesale recipe, a general template of how to think about scientific theories. 

However, structural realism can be bought into also as a retail product, rather than as a wholesale good. In this case, it serves as an interpretive tool, which may or may not fit the theory under consideration. Its application is only justified to the extent to which it fits the case at hand. Such justification must thus be given for each case separately. For instance, it can be argued that since in the presence of quantum entanglement, the total state of a bipartite system does not supervene on the states of the individual subsystems, an ontological commitment to these subsystems must be augmented by the admission of something else into the ontology, such as perhaps the wave function. If instead of choosing such individualism, we approach the interpretation of quantum physics with the structuralist template, then our reconceptualization of individuals in structural terms---and thus our commitment only to the structure of the total system---solves the problem of non-supervenience that the individualist faces. Furthermore, as summarized in \citet[\S3.1]{LadymanRoss}, the structuralist can argue that they escape a form of metaphysical underdetermination that arises from the apparent possibility of permuting indistinguishable elementary particles which the individualist must interpret as resulting in metaphysically distinct situations which remain indiscernible by quantum physics. The individualist must thus accept, it seems, that the individuality of the particles which are indiscernible according to quantum theory transcends what is a complete description of the physical properties of the particles involved, and so conclude that quantum physics must remain incomplete. 

Similarly, as contended in \citet[\S3.2]{LadymanRoss}, a structuralist interpretation of general relativity resolves the impasse between traditional substantivalist and relationalist interpretations of relativistic spacetime and, importantly, removes their respective inadequacies. In a nutshell, (manifold) substantivalists seem to confer to the points of the manifold $M$ of a spacetime $\langle M, g_{ab}\rangle$ with metric $g_{ab}$ an individual existence, which leads to a form of unobservable and physically doubtful form of indeterminism, as exhibited in the so-called `hole argument'. In contrast, relationalists are struck with the problem of the purely `gravitational' degrees of freedom of the metric field: the matter-energy content of the universe as captured by the stress-energy tensor $T_{ab}$ fails to determine the metric $g_{ab}$ of spacetime. As a consequence, it seems impossible to carry through a relationalist reduction of spacetime to matter. Structural realism steps in as a {\em via media} with a reasonable claim that it can deal with both problems. Since it does not commit to the existence of individual spacetime points but only to the spacetime structure as a whole, it evades the pressures from the hole argument. As it posits a relational structure which also includes the gravitational degrees of freedom, it does not fall prey of the relationalist's struggle. 

Although much more could be said about each of the two cases---and much more has of course been said---, they illustrate how structural realism may serve as a useful interpretive template at least for theories in fundamental physics. The retail approach to structural realism differs from the wholesale one that it could have been the case for the former, but not the latter, that a structuralist interpretation would have worked e.g.\ only for quantum physics, but not for general relativity, or vice versa. The central thesis of this section is that causal set theory is naturally amenable to a structuralist reading in the sense of the retail approach, but that this fact has at best mild implications for the wider viability of wholesale structural realism. The rest of this section is primarily concerned with motivating the first part of the thesis, but will remain silent on the latter part. 

An obvious and central task for the structural realist is to articulate the notion of a `structure'. Unfortunately, in much of the literature on structural realism, this important task is neglected. Whenever the notion is explicated, the assumed concept is typically that of a {\em relational structure}. Although category-theoretic notions of structure may ultimately be more general and thus more suitable to fully capture the notion of structure particularly as it is operational in mathematics---another central domain of structural realist ambitions---, the concept of relational structure based on set theory is fully adequate for our purposes.\footnote{Alternative conceptions include, for example, the graph-theoretic notion of structure in \citet{leitlady}, the group-theoretic one in \citet{roberts2011}, and the category-theoretic one in \citet{bain2013}. There is a discussion to be had to what degree these apparently distinct notions may nevertheless be equivalent.}

Roughly,\footnote{For a more rigorous development of the following, see \citet{wut12}.} a {\em relational structure} $\mathcal{S}$ is an ordered pair $\langle O, R\rangle$ of a non-empty set of relations $R$ defined on a non-empty set of relata $O$, the domain of $\mathcal{S}$. An {\em $n$-ary relation} defined on the domain $O$ is a subset of the $n$-fold Cartesian product $O\times \cdots\times O$, where the {\em $n$-fold Cartesian product} of $O$ is defined as the set of all ordered $n$-tuples $\langle x_1, ..., x_n\rangle$ with $x_i\in O$.\footnote{More generally, a Cartesian product (and hence a relation) is defined as of different, in general distinct sets. As this will be irrelevant in what follows, I will ignore this complication.} 

In order to satisfy structuralist demands, the elements of the domain $O$ must not possess any intrinsic nature beyond their structural properties. In other words, they are fully defined by their relational profile, i.e., their set of relations to other elements of $O$. Structural realists sometimes insist that fundamentally, there are no `objects' \citep{french2010}. The best way to make sense of such structuralist claims is precisely to take them to assert that the identity of the elements in the domain is exhausted by their relational profile.

In addition to the characterization of relational structures, we will need a notion of structural identity: what does it mean to say that two structures are the same? We are interested here in {\em physical} structures (rather than in merely mathematical ones), in the sense that the physical systems which exemplify certain structures are ontologically prior to the abstract mathematical structure that may be used to describe them. Thus, under what circumstances can two physical system either in the same physically possible world or in two distinct worlds have the `same' structure? 

The rough idea is that two structures $\mathcal{A}$ and $\mathcal{B}$ are structurally identical in case they have the same relations over their domains $A$ and $B$, which in general will be distinct. This idea is captured by `isomorphisms' between the concerned structures. A {\em homomorphism} from $\mathcal{A}$ to $\mathcal{B}$ is a map $\phi$ from $A$ to $B$ which preserves the relations in the sense that for any $a_i\in A$, if $\langle a_1, ..., a_n\rangle \in R_A$, then $\langle \phi(a_1), ..., \phi(a_n)\rangle \in R_B$. A bijective map $\phi: A\rightarrow B$ is an {\em isomorphism} just in case both $\phi$ and its inverse $\phi^{-1}$ are homomorphisms. Two structures $\mathcal{A}$ and $\mathcal{B}$ are then structurally identical, or {\em isomorphic}, symbolically $\mathcal{A} \simeq \mathcal{B}$, just in case there exists an isomorphism from $A$ to $B$. Finally, two structures $\mathcal{A}$ and $\mathcal{B}$ are {\em automorphic} just in case $\mathcal{A} \simeq \mathcal{B}$ and $A = B$, in which case the corresponding isomorphism is an {\em automorphism}. 

An interpretation of a physical theory $T$ is {\em structuralist} if it asserts that what fundamentally exists according to $T$ just is structural in the sense that it can be fully characterized by automorphism classes of structures. In particular, the elements of the domain of the structures possess no intrinsic nature beyond their function as `carriers' of relational structures identified by their automorphism class. Structural realism pairs such structuralist interpretations with realism. 

With the terminology fixed, it is straightforward to see that causal set theory fits naturally with the concept of a relational structure and is thus amenable to a structuralist interpretation. The domain of the structure just is the set $C$ of basic, featureless events in a causal set, and the single relation fundamentally defined on that domain is the relation of causal precedence $\prec$, which partially orders the set of events. Of course, not all such discrete partial orders are in fact causal sets; as we have seen in \S\ref{ssec:basics}, additional conditions are imposed for these orders to qualify as proper candidates for physical causal sets. However, whatever these additional conditions may be, they do not change the fact that physical causal sets are relational structures $\langle C, \prec\rangle$. 

An important part of a structuralist interpretation, to repeat, is that the elements of $C$ not have any fundamental intrinsic properties apart from their relational profile. As such, it has direct implications for the metaphysics of causal set theory. If basal events are identified merely be their relational profile, then how can two or more events with the same relational profile be distinguished? Connected with the question is the issue of how to interpret highly symmetric structures. 

In order to make this discussion more rigorous, let us introduce the notion of `non-Hegelian' pairs or sets. A {\em non-Hegelian subset} $H\subseteq C$ of events in a causal set $\langle C, \prec\rangle$ is a set of distinct events $x_1,...,x_k$ in $C$ with the same relational profile, i.e., a set $\{ x_1,...,x_k| \forall x_i, x_j, z \in C$ such that $z\neq x_i$ and $z\neq x_j, \neg (x_i\prec x_j)$ and $z\prec x_i \leftrightarrow z\prec x_j$ and $x_i \prec z \leftrightarrow x_j \prec z$, where $i, j= 1,...,k\}$. Any elements of a non-Hegelian set are pairwise unrelated---due to the antisymmetry of $\prec$ they could not have the same relational profile otherwise. 

Although results regarding how generic non-Hegelian sets are in physical causal sets are few and far between, there are reasons to think that they are quite generic in causal sets of sufficient size and with any hope of giving rise to realistic physical structures.\footnote{David Meyer, private communication.} If they do occur, we face a metaphysical conundrum, as those elements will all have an identical relational profile. If the basal events are truly featureless apart from their relational profile, it seems as if these events ought to be considered one and the same: this is Leibniz's {\em principle of the identity of indiscernibles} (PII), according to which whatever cannot be discerned is identical. In other words, two numerically distinct entities must be discernible at least by some of their properties. This raises the well-rehearsed discussion of which properties we quantify over in the PII. In fact, we can distinguish different PIIs with different logical strength, depending on which kinds of properties we consider. 

Whatever the choice, however, given causal set theory's stipulation that the identity of the basal events is exhausted by their relational profile, the only way to keep elements of non-Hegelian sets distinct is by endowing them with a `primitive thisness' or `haecceities'. The idea here is that it is part of the essence of a basal event to be that particular event and no other. Thus, two elements of a non-Hegelian set are distinct by virtue of possessing a distinct essential haecceities. Many philosophers of physics feel queasy in the face of haecceities, as these are essential and fundamental yet completely intangible properties. From a philosophical perspective, many would thus prefer not to have to rely on haecceities and so to identify the elements of non-Hegelian sets. Physicists often have practical reasons to do the same, for instance because the matrix encoding the causal structure is otherwise degenerate and so not invertible or because non-standard set theory would have to be used instead of the standard one.

However, at least at the level of general relativity, many of the most important spacetimes have a high degree of internal symmetry, such as spatial homogeneity or isotropy or invariance under time translations. Given that classes of events in those cases cannot be discerned by means of fundamental physical properties and yet it would be absurd to identify for instance all events on a spacelike hypersurface of Friedmann-Lema\^{\i}tre-Robertson-Walker spacetimes---or all events of Minkowski spacetime---, one might worry that more tolerance for non-Hegelian sets in causal set theory is called for. 

Perhaps such worry is unnecessary. Consider Minkowski spacetime. In order to give rise to Lorentz symmetry, the fundamental causal set cannot be too regular or symmetric; if it were, then there would be a physically privileged foliation at the emergent level \citep{bomeal09}. Thus, in order to give rise to a highly symmetric continuum spacetime, the fundamental discrete structure cannot be too symmetric. This mechanism ascertains that non-Hegelian sets cannot play too much of a role for physically relevant models of causal set theory. 

In sum, the consideration of structural realism in the context of causal set theory is fruitful in both directions. On the one hand, structural realism suggests a very natural interpretation of causal set theory and arguably helps clarifying the metaphysics of causal sets in ways that may have implications in the technical formulation of the theory. On the other hand, causal set theory offers a particularly elegant exemplar for the structural realist to develop and articulate their position.

\section{Conclusions}
\label{sec:conc}

Causal set theory offers a very rich field for philosophical study, due to different aspects: the suggestion that causal relations may be more fundamental than temporal ones, the functionalist emergence particularly of all spatial structures, the proposed (still classical) dynamics evoking a relativistically kosher, `asynchronous' form of becoming, and its straightforwardly structuralist interpretation. At least the first three remain unsettled as their discussion continues. I hope to have presented the state of these debates in an engaging way and to have shown their fruitful connections to larger issues in the foundations of physics, the philosophy of time, and the metaphysics of science.

\bibliographystyle{plainnat}
\bibliography{biblio}

\begin{thebibliography}{58}
\providecommand{\natexlab}[1]{#1}
\providecommand{\url}[1]{\texttt{#1}}
\expandafter\ifx\csname urlstyle\endcsname\relax
  \providecommand{\doi}[1]{doi: #1}\else
  \providecommand{\doi}{doi: \begingroup \urlstyle{rm}\Url}\fi

\bibitem[Arageorgis(2016)]{Arageorgis12}
Aristidis Arageorgis.
\newblock Spacetime as a causal set: universe as a growing block?
\newblock \emph{Belgrade Philosophical Annual}, 29:\penalty0 33--55, 2016.

\bibitem[Bain(2013)]{bain2013}
Jonathan Bain.
\newblock Category-theoretic structure and radical ontic structural realism.
\newblock \emph{Synthese}, 190:\penalty0 1621--1635, 2013.

\bibitem[Baron and Le~Bihan(2022)]{lebbar22}
Sam Baron and Baptiste Le~Bihan.
\newblock Quantum gravity and mereology: not so simple.
\newblock \emph{The Philosophical Quarterly}, 72:\penalty0 19--40, 2022.

\bibitem[Baron and Le~Bihan(forthcoming)]{baronlebihan2023}
Sam Baron and Baptiste Le~Bihan.
\newblock Causal theories of spacetime.
\newblock \emph{No\^us}, forthcoming.

\bibitem[Bell(1987)]{bel87}
John~S Bell.
\newblock \emph{Speakable and Unspeakable in Quantum Mechanics}.
\newblock Cambridge University Press, Cambridge, 1987.

\bibitem[Bombelli et~al.(1987)Bombelli, Lee, Meyer, and Sorkin]{bomeal87}
Luca Bombelli, Joohan Lee, David Meyer, and Rafael Sorkin.
\newblock Spacetime as a causal set.
\newblock \emph{Physical Review Letters}, 59:\penalty0 521--524, 1987.

\bibitem[Bombelli et~al.(2009)Bombelli, Henson, and Sorkin]{bomeal09}
Luca Bombelli, Joe Henson, and Rafael~D Sorkin.
\newblock Discreteness without symmetry breaking: a theorem.
\newblock \emph{Modern Physics Letters A}, 24:\penalty0 2579--2587, 2009.

\bibitem[Brightwell(1997)]{bri97}
Graham Brightwell.
\newblock Partial orders.
\newblock In Lowell~W Beineke and Robin~J Wilson, editors, \emph{Graph
  Connections: Relationships between Graph Theory and other Areas of
  Mathematics}, pages 52--69. Clarendon Press, Oxford, 1997.

\bibitem[Butterfield(2007)]{Butterfield2007}
Jeremy Butterfield.
\newblock Stochastic {Einstein} locality revisited.
\newblock \emph{British Journal for the Philosophy of Science}, 58:\penalty0
  805--867, 2007.

\bibitem[Callender(2000)]{cal00}
Craig Callender.
\newblock Shedding light on time.
\newblock \emph{Philosophy of Science}, 67:\penalty0 S587--S599, 2000.

\bibitem[Clifton and Hogarth(1995)]{CliftonHogarth}
Rob Clifton and Mark Hogarth.
\newblock The definability of objective becoming in {Minkowski} spacetime.
\newblock \emph{Synthese}, 103:\penalty0 355--287, 1995.

\bibitem[Dowker(2003)]{Dowker2003}
Fay Dowker.
\newblock Real time.
\newblock \emph{New Scientist}, pages 36--39, 4 October 2003.

\bibitem[Dowker(2013)]{Dowker2013}
Fay Dowker.
\newblock Introduction to causal sets and their phenomenology.
\newblock \emph{General Relativity and Gravitation}, 45:\penalty0 1651--1667,
  2013.

\bibitem[Dowker(2014)]{dowker14}
Fay Dowker.
\newblock The birth of spacetime atoms as the passage of time.
\newblock \emph{Annals of the New York Academy of Sciences}, 1326:\penalty0
  18--25, 2014.

\bibitem[Dowker(2020)]{dowker20}
Fay Dowker.
\newblock Being and becoming on the road to quantum gravity: or, the birth of a
  baby is not a baby.
\newblock In Nick Huggett, Keizo Matsubara, and Christian W\"uthrich, editors,
  \emph{Beyond Spacetime: The Foundations of Quantum Gravity}, pages 133--142.
  Cambridge University Press, Cambridge, 2020.

\bibitem[Earman(2008)]{Earman2008}
John Earman.
\newblock Reassessing the prospects for a growing block model of the universe.
\newblock \emph{International Studies in the Philosophy of Science},
  22:\penalty0 135--164, 2008.

\bibitem[Emery et~al.(2020)Emery, Markosian, and Sullivan]{emery20}
Nina Emery, Ned Markosian, and Meghan Sullivan.
\newblock Time.
\newblock \emph{Stanford Encyclopedia of Philosophy}, 2020.
\newblock URL \url{https://plato.stanford.edu/entries/time/}.

\bibitem[Fine(2005)]{fine2005}
Kit Fine.
\newblock Tense and reality.
\newblock In Kit Fine, editor, \emph{Modality and Tense: Philosophical Papers},
  pages 259--320. Oxford University Press, Oxford, 2005.

\bibitem[French(2010)]{french2010}
Steven French.
\newblock The interdependence of structure, objects, and dependence.
\newblock \emph{Synthese}, 175 (supplement):\penalty0 89--109, 2010.

\bibitem[Huggett(2014)]{hug14}
Nick Huggett.
\newblock Skeptical notes on a physics of passage.
\newblock \emph{Annals of the New York Academy of Sciences}, 1326:\penalty0
  9--17, 2014.

\bibitem[Huggett and W\"uthrich(2013)]{hugwut13}
Nick Huggett and Christian W\"uthrich.
\newblock Emergent spacetime and empirical (in)coherence.
\newblock \emph{Studies in History and Philosophy of Modern Physics},
  44:\penalty0 276--285, 2013.

\bibitem[Huggett and W\"uthrich(Under contract)]{hugwut}
Nick Huggett and Christian W\"uthrich.
\newblock \emph{Out of Nowhere: The Emergence of Spacetime in Quantum Theories
  of Gravity}.
\newblock Oxford University Press, Oxford, Under contract.

\bibitem[Kleitman and Rothschild(1975)]{klerot75}
Daniel~J Kleitman and Bruce~L Rothschild.
\newblock Asymptotic enumeration of partial orders on a finite set.
\newblock \emph{Transactions of the American Mathematical Society},
  205:\penalty0 205--220, 1975.

\bibitem[Ladyman and Ross(2007)]{LadymanRoss}
James Ladyman and Don Ross.
\newblock \emph{Every Thing Must Go: Metaphysics Naturalized}.
\newblock Oxford University Press, Oxford, 2007.

\bibitem[Lam and W\"uthrich(2018)]{lamwut18}
Vincent Lam and Christian W\"uthrich.
\newblock Spacetime is as spacetime does.
\newblock \emph{Studies in History and Philosophy of Modern Physics},
  64:\penalty0 39--51, 2018.

\bibitem[Lam and W\"uthrich(2021)]{lamwut20}
Vincent Lam and Christian W\"uthrich.
\newblock Spacetime functionalism from a realist perspective.
\newblock \emph{Synthese}, 199:\penalty0 335--353, 2021.

\bibitem[Le~Bihan(2021)]{le2021spacetime}
Baptiste Le~Bihan.
\newblock Spacetime emergence in quantum gravity: Functionalism and the hard
  problem.
\newblock \emph{Synthese}, 199\penalty0 (2):\penalty0 371--393, 2021.

\bibitem[Le~Bihan and Linnemann(2019)]{leblin19}
Baptiste Le~Bihan and Niels Linnemann.
\newblock Have we lost spacetime on the way? {Narrowing} the gap between
  general relativity and quantum gravity.
\newblock \emph{Studies in History and Philosophy of Modern Physics},
  65:\penalty0 112--121, 2019.

\bibitem[Leitgeb and Ladyman(2008)]{leitlady}
Hannes Leitgeb and James Ladyman.
\newblock Criteria of identity and structuralist ontology.
\newblock \emph{Philosophia Mathematica}, 16:\penalty0 388--396, 2008.

\bibitem[Levin(2021)]{lev21}
Janet Levin.
\newblock Functionalism.
\newblock \emph{Stanford Encyclopedia of Philosophy}, 2021.
\newblock URL
  \url{https://plato.stanford.edu/archives/win2021/entries/functionalism/}.

\bibitem[Lombard(1999)]{lom99}
Lawrence Lombard.
\newblock On the alleged incompatibility of presentism and temporal parts.
\newblock \emph{Philosophia}, 27:\penalty0 253--260, 1999.

\bibitem[Major et~al.(2006)Major, Rideout, and Surya]{Major2006}
Seth Major, David Rideout, and Sumati Surya.
\newblock {Spatial hypersurfaces in causal set cosmology}.
\newblock \emph{Classical and Quantum Gravity}, 23:\penalty0 4743--4752, 2006.

\bibitem[Major et~al.(2007)Major, Rideout, and Surya]{Major2007}
Seth Major, David Rideout, and Sumati Surya.
\newblock {On recovering continuum topology from a causal set}.
\newblock \emph{Journal of Mathematical Physics}, 48:\penalty0 032501, 2007.

\bibitem[Malament(1977)]{mal77}
David~B Malament.
\newblock The class of continuous timelike curves determines the topology of
  spacetime.
\newblock \emph{Journal of Mathematical Physics}, 18:\penalty0 1399--1404,
  1977.

\bibitem[Meyer(2005)]{mey05}
Ulrich Meyer.
\newblock The presentist's dilemma.
\newblock \emph{Philosophical Studies}, 122:\penalty0 213--225, 2005.

\bibitem[Ney(2015)]{ney15}
Alyssa Ney.
\newblock Fundamental physical ontologies and the constraint of empirical
  coherence.
\newblock \emph{Synthese}, 192:\penalty0 3105--3124, 2015.

\bibitem[Putnam(1967)]{put67}
Hilary Putnam.
\newblock Time and physical geometry.
\newblock \emph{Journal of Philosophy}, 64:\penalty0 240--247, 1967.

\bibitem[Rideout and Wallden(2009{\natexlab{a}})]{ridwal09a}
David Rideout and Petros Wallden.
\newblock Emergence of spatial structure from causal sets.
\newblock \emph{Journal of Physics: Conference Series}, 174:\penalty0 012017,
  2009{\natexlab{a}}.

\bibitem[Rideout and Wallden(2009{\natexlab{b}})]{ridwal09b}
David Rideout and Petros Wallden.
\newblock Spacelike distance from discrete causal order.
\newblock \emph{Classical and Quantum Gravity}, 26:\penalty0 155013,
  2009{\natexlab{b}}.

\bibitem[Rideout and Sorkin(1999)]{Rideout1999}
David~P. Rideout and Rafael~D. Sorkin.
\newblock Classical sequential growth dynamics for causal sets.
\newblock \emph{Physical Review D}, 61:\penalty0 024002, 1999.

\bibitem[Rietdijk(1966)]{rie66}
C~W Rietdijk.
\newblock A rigorous proof of determinsim derived from the special theory of
  relativity.
\newblock \emph{Philosophy of Science}, 33:\penalty0 341--344, 1966.

\bibitem[Roberts(2011)]{roberts2011}
Bryan Roberts.
\newblock Group structural realism.
\newblock \emph{British Journal for the Philosophy of Science}, 62:\penalty0
  47--69, 2011.

\bibitem[Saunders(2002)]{sau02}
Simon Saunders.
\newblock How relativity contradicts presentism.
\newblock In Craig Callender, editor, \emph{Time, Reality and Experience},
  pages 277--292. Cambridge University Press, Cambridge, 2002.

\bibitem[Savitt(2006)]{sav06}
Steven Savitt.
\newblock Presentism and eternalism in perspective.
\newblock In Dennis Dieks, editor, \emph{The Ontology of Spacetime}, pages
  111--127. Elsevier, Amsterdam, 2006.

\bibitem[Sklar(1981)]{skl81}
Lawrence Sklar.
\newblock Time, reality, and relativity.
\newblock In Richard Healey, editor, \emph{Reduction, Time, and Reality}, pages
  129--142. Cambridge University Press, Cambridge, 1981.

\bibitem[Sorkin(2006)]{Sorkin2006}
Rafael~D Sorkin.
\newblock Geometry from order: causal sets.
\newblock \emph{Einstein Online}, 2:\penalty0 1007, 2006.

\bibitem[Sorkin(2007)]{Sorkin2007}
Rafael~D Sorkin.
\newblock Relativity theory does not imply that the future already exists: A
  counterexample.
\newblock In Vesselin Petkov, editor, \emph{Relativity and the Dimensionality
  of the World}, pages 153--161. Springer, Dordrecht, 2007.

\bibitem[Sorkin(2009)]{Sorkin2009b}
Rafael~D. Sorkin.
\newblock Does locality fail at intermediate length-scales?
\newblock In Daniele Oriti, editor, \emph{Approaches to Quantum Gravity: Toward
  a New Understanding of Space, Time and Matter}, chapter~3, pages 26--43.
  Cambridge University Press, Cambridge, 2009.

\bibitem[Stein(1991)]{Stein1991}
Howard Stein.
\newblock On relativity theory and openness of the future.
\newblock \emph{Philosophy of Science}, 58:\penalty0 147--167, 1991.

\bibitem[Surya(2019)]{Surya2019}
Sumati Surya.
\newblock The causal set approach to quantum gravity.
\newblock \emph{Living Reviews in Relativity}, 22\penalty0 (5), 2019.

\bibitem[Worrall(1989)]{Worrall1989}
John Worrall.
\newblock Structural realism: the best of both worlds?
\newblock \emph{Dialectica}, 43:\penalty0 99--124, 1989.

\bibitem[W\"uthrich(2012{\natexlab{a}})]{wut12}
Christian W\"uthrich.
\newblock The structure of causal sets.
\newblock \emph{Journal for General Philosophy of Science}, 43:\penalty0
  223--241, 2012{\natexlab{a}}.

\bibitem[W\"uthrich(2012{\natexlab{b}})]{wut12c}
Christian W\"uthrich.
\newblock Demarcating presentism.
\newblock In Henk de~Regt, Samir Okasha, and Stephan Hartmann, editors,
  \emph{EPSA Philosophy of Science: Amsterdam 2009}, pages 439--448. Springer,
  Dordrecht, 2012{\natexlab{b}}.

\bibitem[W\"uthrich(2013)]{wut13}
Christian W\"uthrich.
\newblock The fate of presentism in modern physics.
\newblock In Roberto Ciuni, Kristie Miller, and Giuliano Torrengo, editors,
  \emph{New Papers on the Present: Focus on Presentism}, pages 91--131.
  Philosophia Verlag, Munich, 2013.

\bibitem[W\"uthrich(2021)]{wut21}
Christian W\"uthrich.
\newblock Time travelling in emergent spacetime.
\newblock In Judit Madar\'asz and Gergely Sz\'ekely, editors, \emph{Hajnal
  Andr\'eka and Istv\'an N\'emeti on the Unity of Science: From Computing to
  Relativity Theory Through Algebraic Logic}, number~19 in Outstanding
  Contributions to Logic, pages 453--474. Springer, Cham, 2021.

\bibitem[W\"uthrich and Callender(2017)]{wutcal15}
Christian W\"uthrich and Craig Callender.
\newblock What becomes of a causal set.
\newblock \emph{British Journal for the Philosophy of Science}, 68:\penalty0
  907--925, 2017.

\bibitem[Yates(2020)]{Yat:19}
David Yates.
\newblock Thinking about spacetime.
\newblock In Christian W\"uthrich, Baptiste~Le Bihan, and Nick Huggett,
  editors, \emph{Philosophy Beyond Spacetime: Implications from Quantum
  Gravity}, pages 129--153. Oxford University Press, Oxford, 2020.

\bibitem[Zalel(forthcoming)]{Zalel23}
Stav Zalel.
\newblock Covariant growth dynamics.
\newblock In Cosimo Bambi, Leonardo Modesto, and Ilya Shapiro, editors,
  \emph{Handbook of Quantum Gravity}. Springer Nature, Singapore, forthcoming.

\end{thebibliography}

\end{document}